\documentclass[cameraready]{Interspeech}
\title{Endpoint Anticipation for Low-Latency Spoken Dialogue}

\author[affiliation={1}]{Sathvik}{Udupa}
\author[affiliation={2}]{Shinji}{Watanabe}
\author[affiliation={1}]{Petr}{Schwarz}
\author[affiliation={1}]{Jan}{Cernocky}
\address{
    $^1$ Brno University of Technology, Czechia \\
    $^2$ Carnegie Mellon University, United States
}

\email{\{udupa, schwarzp, cernocky\}@fit.vut.cz, shinjiw@ieee.org}

\usepackage{arydshln}

\usepackage{comment}
\usepackage{cite}
\usepackage{multirow}
\usepackage{makecell}
\usepackage{xcolor}
\definecolor{bluevlgt}{RGB}{219,234,254}
\definecolor{bluelgt}{RGB}{147,197,253}
\definecolor{bluemed}{RGB}{59,130,246}
\definecolor{bluedark}{RGB}{29,78,216}
\definecolor{redlgt}{RGB}{254,202,202}
\begin{document}
\maketitle

\begin{abstract}
While low-latency interaction is critical for spoken dialogue, cascaded architectures are often bottlenecked by reactive turn-completion detection. We propose \textit{Endpoint Anticipation}, shifting from reactive detection to proactive forecasting of end-of-turn signals. Our speech-based model anticipates endpoints up to 2.56 seconds in advance, enabling speculative execution of LLM and TTS pipelines on partial context. We introduce metrics to quantify the trade-off between realized latency reduction and computational redundancy. Evaluation across conversational and task-oriented datasets shows our model consistently outperforms competitive VAP-based baselines. Integration with the \textit{Unmute} framework demonstrates a 505 ms average latency reduction with a 28.4\% increase in speculative computation, effectively masking sequential bottlenecks to enable complex reasoning in real-time speech-to-speech interaction.
\end{abstract}

% TODO - 
% VAP
%   Re-run VAP ETA 8 pm - 640 ms, 1280 ms
%   Score VAP
%   Change fig 1
%   Change 4.2
%   Change 5.1

% conv vs spoken - OK

% STL vs MHL
% redraw fig 3
%  rewrite 5.3

% unmute
% Fix generate
% fix table 2
% fix 5.4

% fix all plot colours and legend locs

% fix refrences

% See if a figure can be added to explain metrics

\section{Introduction}
Real-time spoken dialogue systems \cite{arora2025landscape} have seen significant growth, driven by advances in large language models (LLMs). These systems \cite{defossez2024moshi, hu2025salm, roy2026personaplexvoicerolecontrol, likhomanenko2025chipchat, yu2025salmonn, wang2025end, zeng2024glm, zufle2026f} aim to process audio with low latency to perform complex tasks, often utilizing streaming speech inputs paired with the reasoning capabilities of LLMs to generate responses.

While end-to-end training offers a path to low latency \cite{defossez2024moshi}, many competitive frameworks such as Unmute\footnote{\url{https://github.com/kyutai-labs/unmute}} rely on cascaded architectures that control turn-taking via an \textit{endpointer} \cite{chang17_interspeech, inoue2024real, udupa2025streaming, li2025easy}. In these systems, detecting an endpoint triggers a sequential pipeline --- completing automatic speech recognition (ASR), LLM generation, and Text-to-Speech (TTS) --- which imposes a theoretical lower bound on the Time-to-First-Audio (TTFA). Consequently, while humans respond within $\sim$250 ms \cite{stivers2009universals}, modular setups like ChipChat, Unmute, and Pipecat\footnote{\url{https://github.com/pipecat-ai/pipecat}} exhibit latencies close to 1--2 seconds \cite{likhomanenko2025chipchat, kuroki2025kame}. This gap persists because current systems are reactive, whereas human listeners actively anticipate turn completions using linguistic and prosodic cues to minimize gaps \cite{levinson2015timing}. This cascaded bottleneck, where generation cannot begin until speech ends, creates a significant barrier for integrating further processing such as dialogue management, tool-use, or reasoning.

To overcome this, we introduce \textit{endpoint anticipation}, a framework that forecasts EOT signals before turn completion. Unlike standard endpointing, this early forecasting allows the system to initiate LLM and TTS processing during the user's ongoing speech. By generating initial hypotheses and pre-fetching audio frames during the user's speech, we ``pipeline" the generation process to significantly mask system latency.

To demonstrate viability, we integrate our model into the Unmute framework using speculative execution. An anticipated EOT triggers the system to initiate the LLM-TTS pipeline in the background. If the user continues speaking, this constitutes a premature trigger. The system subsequently discards the speculative response at the next anticipated signal. Upon a confirmed endpoint, we complete the pre-buffered prefix with the pending transcript and continue generation seamlessly from the speculative tokens. This mechanism introduces a fundamental trade-off between latency reduction and the computational cost of discarded generations. Consequently, we propose a comprehensive set of metrics to quantify this balance. These metrics strictly distinguish between successful anticipation horizons and premature prediction rates.

The contributions of this work are as follows:
\begin{enumerate}
\item We propose a speech-based \textbf{Endpoint Anticipation} task and model designed for low-latency spoken dialogue systems.
\item We define a set of metrics to quantify the trade-off between Realized Anticipation (the actual latency reduction provided within the target window) and Premature Anticipation (the resulting downstream computational redundancy due to predictions made before the valid horizon).
\item We evaluate the framework across various anticipation targets ranging from 320 ms to 2560 ms.
\item We open-source our implementation and provide a reference integration with the Unmute full-duplex framework.\footnote{https://github.com/bloodraven66/EndpointAnticipation}
\end{enumerate}
\section{Related Work}

\textbf{Early End-of-Utterance Prediction.} Several approaches utilize ASR to predict End-of-Utterance (EOU) tokens ahead of time. Sakuma et al. \cite{sakuma2023response, sakuma2023improving} propose a two-stage method: first generating a text hypothesis from streaming ASR, followed by a language model that predicts future EOU tokens. Chang et al. \cite{chang2020low} similarly exploit early endpoint signals to prefetch downstream responses, controlling the trigger via a confidence threshold; our method differs in that we parameterize prefetching through a fixed anticipation horizon rather than a decision boundary, decoupling it from ASR confidence entirely. Similarly, Zink et al. \cite{zink2024predictive} leverage the decoder of an encoder-decoder ASR model for early EOU prediction.  Unlike these text-dependent approaches, our work introduces a speech-only forecasting model that operates directly on the acoustic signal, entirely bypassing the ASR bottleneck.

\noindent\textbf{Voice Activity Projection (VAP).} VAP \cite{ekstedt2022voice, inoue-etal-2024-multilingual, inoue2024real} projects the future voice activity of two speakers to learn general turn-taking abilities, introducing a \textsc{Turn-Shift} task to predict near-future speaker changes. While VAP focuses on generalized dialogue states over discrete bins, our work directly targets continuous, fixed-horizon endpoint forecasting. Consequently, we adapt VAP's projected future probabilities to serve as our primary baseline for comparative evaluation.

\noindent\textbf{Computation-Intensive SDS.} Recent spoken dialogue systems increasingly incorporate complex, real-time processing. Recent works \cite{shih2025can, arora25_interspeech} introduce chain-of-thought reasoning for end-to-end SDS, while KAME \cite{kuroki2025kame} uses a secondary LLM for reasoning over partial transcripts, and Stream RAG \cite{arora2025stream} enables low-latency audio tool-calling. While end-to-end architectures natively accommodate such computational demands, cascaded systems remain constrained by sequential latency bottlenecks. Our framework mitigates this limitation, enabling modular pipelines to integrate complex processing while preserving low-latency interactions.

\section{Proposed approach}
In this section, we describe the model backbone and introduce the modeling framework for endpoint anticipation.

\subsection{Dual-stream audio representation}
\label{ref:prop_dual_stream}
Similar to \cite{ekstedt2022voice, udupa2025streaming}, we process User ($u$) and System ($s$) audio streams to provide interaction context. Let $t$ denote the timestamp of the current frame. Let $\mathbf{X}_{\le t}^{(u)}$ and $\mathbf{X}_{\le t}^{(s)}$ represent the sequences of audio features extracted up to frame $t$ from their respective channels. We employ two independent streaming Transformer encoders, $\mathcal{T}_{u}$ and $\mathcal{T}_{s}$, to process each stream:
\begin{equation}
\mathbf{Z}^{(u)}_{\le t} = \mathcal{T}_{u}(\mathbf{X}^{(u)}_{\le t}), \quad \mathbf{Z}^{(s)}_{\le t} = \mathcal{T}_{s}(\mathbf{X}^{(s)}_{\le t})
\end{equation}

The resulting latent representations are concatenated along the feature dimension to form a unified context vector $\mathbf{Z}_{\le t} = [\mathbf{Z}^{(u)}_{\le t}; \mathbf{Z}^{(s)}_{\le t}]$. This fused representation is then passed to the prediction heads described below. This ensures the model can learn to anticipate endpoints in the presence of turn-taking conditions such as backchannels and interruptions.

\subsection{Endpoint Anticipation (EPA)}
We formulate \textit{Endpoint Anticipation} (EPA) as a set of independent binary classification tasks. Let $\mathcal{H} = \{320, 640, \dots, 2560\}$ be the set of anticipation horizons in milliseconds. For each horizon $h \in \mathcal{H}$, the binary target $y_t^{(h)}$ at time $t$ is defined as:
\begin{equation}
    y_t^{(h)} = \begin{cases} 
       1 & \text{if } 0 \le t_{\mathrm{EOT}} - t \le h, \\
       0 & \text{otherwise}
    \end{cases}
    \label{eq:target_def}
\end{equation}
where $t_{\mathrm{EOT}}$ is the timestamp of the user's turn completion. We evaluate predictors for all $h \in \mathcal{H}$. At our feature frame rate of 12.5 Hz (80 ms period), the smallest horizon $h = 320$ corresponds to an anticipation window of exactly 4 frames.

During inference, the model estimates the probability $p_t^{(h)}$ of an upcoming endpoint. The binary anticipation decision $\hat{y}_t^{(h)}$ is triggered using a predefined threshold $\theta$:
\begin{equation}
\hat{y}_t^{(h)} = \begin{cases}
1 & \text{if } p_t^{(h)} \ge \theta, \\
0 & \text{otherwise}
\end{cases}
\label{eq:inference_def}
\end{equation}
This threshold dictates the operating point. It allows tuning the balance between latency reduction and premature predictions. During deployment, the first frame where $\hat{y}_t^{(h)} = 1$ acts as the trigger for the speculative pipeline.

\subsection{Model Architectures}
\label{sec:model_arch}

We investigate two modeling strategies to predict the set of anticipation windows $\{y_t^{(h)}\}_{h \in \mathcal{H}}$ defined in Eq. \ref{eq:target_def}.

\noindent\textbf{EPA-S (Single-Target Learning):} We train $|\mathcal{H}|$ independent models, one for each horizon. For a specific horizon $h \in \mathcal{H}$, the dedicated model extracts the concatenated context vector $\mathbf{Z}_{\le t}$ and estimates the endpoint probability as:
\begin{equation}
    {p}_t^{(h)} = \sigma(\mathbf{W} \cdot \mathbf{Z}_{\le t} + b)
\end{equation}
where $\sigma$ is the sigmoid function, and $\mathbf{W}$ and $b$ are the parameters of the projection head. While flexible, the computational and memory cost scales linearly with $|\mathcal{H}|$. Note that, unlike VAP \cite{ekstedt2022voice}, the loss is computed only for one primary speaker, and the features from other speaker act as conversation context.

\noindent\textbf{EPA-M (Multi-target Learning):} To improve efficiency, we use Multi-Task Learning with a shared dual-stream backbone. The model branches only at the final layer to produce predictions for all horizons simultaneously:
\begin{equation}
    {p}_t^{(h)} = \sigma(\mathbf{W}_h \cdot \mathbf{Z}_{\le t} + b_h), \quad \forall h \in \mathcal{H}
\end{equation}
where $\mathbf{Z}_{\le t}$ is the shared latent representation. This setup learns generalized turn-completion features while maintaining horizon-specific decision boundaries via the heads $\{\mathbf{W}_h, b_h\}$.

\section{Experimental setup}

\subsection{Dataset}
We train and evaluate on SpokenWOZ \cite{si2023spokenwoz} (8 kHz, task-oriented) and Switchboard \cite{godfrey1992switchboard} (8 kHz, conversational), modeling the \textit{User} and \textit{Speaker A} streams as primary speakers, respectively. To ensure precise endpoint supervision, we refine raw turn boundaries using Silero VAD \cite{SileroVAD} to strip trailing silence. To prevent premature predictions at speech onset and exclude backchannels\footnote{We mask the loss during primary speaker backchannels, as they lack sufficient context for forecasting turn completion.}, we mask the loss for turns shorter than 2 seconds (also requiring a min. of 3 words for Switchboard).

\subsection{Baseline}
We use the pretrained Voice Activity Projection (VAP)\footnote{\url{https://github.com/ErikEkstedt/VAP}} \cite{ekstedt2022voice} as our baseline. Because VAP outputs 50 Hz turn-taking probabilities over discrete temporal bins, we adapt it to our continuous 12.5 Hz horizon-based evaluation. Specifically, we extract the non-active speaker's probability from the \textit{p\_future} distribution and downsample it to 12.5 Hz via 4-frame mean pooling. To establish the most competitive baseline, we ran inference on different \textit{p\_future} bin combinations and present the best results; we apply our evaluation threshold to the pooled probabilities of bins [0--1] (0--600 ms) for the 640 ms anticipation window, and bins [1--2] (200--1200 ms) for the 1280 ms window.

\subsection{Metrics}
While prior works typically rely on aggregate precision and recall \cite{ekstedt2022voice}, we introduce four metrics explicitly designed to quantify the practical trade-offs between latency reduction (successful anticipation) and wasted computation (premature predictions) in real-time systems:

\begin{itemize}
    \item \textbf{Median Realized Anticipation (MRA):} Measures the actual latency savings. For turns with a successful prediction inside the valid anticipation window $[t_{\mathrm{EOT}} - h, t_{\mathrm{EOT}}]$, we calculate the duration between the first valid prediction $t_{\mathrm{pred}}$ and the true endpoint $t_{\mathrm{EOT}}$. We report the median of $t_{\mathrm{EOT}} - t_{\mathrm{pred}}$ across the test set.
    \item \textbf{Premature Anticipation Rate (PAR):} Assesses turn-level stability by tracking activations that occur before the valid anticipation window (i.e., $t_{\mathrm{pred}} < t_{\mathrm{EOT}} - h$ for a given $h \in \mathcal{H}$). These early triggers result in wasteful downstream computation. PAR is the percentage of turns containing at least one such premature activation.  
    \item \textbf{Expected Redundant Computation (ERC):} Because longer turns present more opportunities for early triggers, PAR can be biased by turn duration. We define ERC as the ratio of actual premature anticipations to the maximum possible anticipations for a turn of length $T$, approximated by $\lceil (T - h) / h \rceil$. Reported as a mean percentage across all turns, this metric quantifies the expected proportion of discarded speculative computation. 
    \item \textbf{Horizon Entry Accuracy (HEA):} Evaluates the model's temporal precision in triggering exactly at the target horizon boundary $t = t_{\mathrm{EOT}} - h$. We frame this as a binary classification task where predictions within a tight two-frame collar $\{t, t+1\}$ are true positives\footnote{The maximum anticipation provided by the selected VAP bins falls within this collar, ensuring a fair baseline comparison.}, and predictions in the subsequent $[t+2, t_{\mathrm{EOT}}]$ range are false positives. This ensures the model reliably matches the requested anticipation.
\end{itemize}
Metrics are computed only for turns exceeding the horizon window ($T > h$). We present the results as latency versus early-trigger curves at different prediction probability thresholds.

\subsection{Training Setup}
\textbf{Feature Extraction:} 
We use Mimi neural codec \cite{defossez2024moshi} (using the first 8 codebooks) as the feature backbone based on preliminary experiments. Input audio is upsampled to 24 kHz and features are extracted at 12.5 Hz with zero lookahead. We freeze the backbone parameters, enabling a modular design where the same feature extractor can serve many downstream tasks.

\noindent\textbf{Model Configuration:} The model is a 25M parameter streaming Transformer \cite{vaswani2017attention}. It comprises a 6-layer encoder with 4 attention heads and a feed-forward dimension of 1024. For long-form streaming, we apply RoPE \cite{su2024roformer} and causal masking with a fixed 250-frame left context.

\noindent\textbf{Optimization:} All proposed models are trained on both the SpokenWOZ and Switchboard datasets. We use a learning rate of $3 \times 10^{-4}$ and a batch size of 16. During training, we sample fixed-length segments of 500 frames (40 seconds). To address class imbalance, we apply a 10:1 weighted loss between the positive (horizon, $y_t^{(h)} = 1$) and negative (non-horizon, $y_t^{(h)} = 0$) classes. We mask loss for turns shorter than 2 seconds to prevent early predictions. We track the mean accuracy across horizon and non-horizon frames and apply early stopping with a patience of 6 epochs if no improvement is observed.

\subsection{Speech-to-Speech Integration}
\label{sec:speech_speech_integration}
We demonstrate the utility of Endpoint Anticipation by integrating it into the \textit{Unmute} framework. This section outlines the baseline architecture and our proposed speculative execution strategy to validate latency reduction in real-world settings.

\subsubsection{Unmute Framework}
The Unmute system is a modular, speech-to-speech architecture designed for low-latency interaction, utilizing a pipeline of streaming ASR, TTS \cite{zeghidour2025streaming} and LLM. It uses a semantic VAD endpointer to determine when to trigger system responses. The backend is a Rust-based WebSocket server, while the Python client manages turn-taking logic and interruption handling. We use Gemma 3 4B\footnote{\url{https://huggingface.co/google/gemma-3-4b-it}} as the LLM, hosted using vLLM \cite{kwon2023efficient}.

\begin{table}[t]
    \centering
    \resizebox{\textwidth/2}{!}{
    \begin{tabular}{c|c|c|c|c|c}
    \hline
         Model & $h$ &  MRA (ms) $\uparrow$ & HEA (\%)  $\uparrow$ & PAR (\%) $\downarrow$ & ERC $\downarrow$  \\
         \hline
         VAP & 640 & 160 & 19.2 & 68.3 & \textcolor{blue}{34.5} \\
         EPA-S & 640 & \textbf{640} & 66.3 & 66.5 & \textcolor{blue}{33.9} \\
         EPA-M & 640 & \textbf{640} & \textbf{67.0} & \textbf{66.2} & \textcolor{blue}{33.8} \\
         \hdashline
         VAP & 1280 & 320 & 20.8 & \textbf{51.0} & \textcolor{blue}{33.8} \\
         EPA-S & 1280 & \textbf{1200} & \textbf{50.3} & 53.9 & \textcolor{blue}{33.7} \\
         EPA-M & 1280 & 1120 & 49.7 & 52.8 & \textcolor{blue}{33.2} \\
         \hline
         VAP & 1280 & 80 & 7.2 & \textbf{28.8} & \textcolor{purple}{15.4} \\
         EPA-S & 1280 & \textbf{480} & 22.0 & 34.4 & \textcolor{purple}{15.1} \\ 
         EPA-M & 1280 & \textbf{480} & \textbf{22.1} & 34.3 & \textcolor{purple}{15.1} \\
         \hline
    \end{tabular}}
    \caption{Metrics for two specific operating points - \textcolor{blue}{ERC $\approx$ 33 \%} and \textcolor{purple}{ERC $\approx$ 15 \%}. VAP baseline, along with EPA-S and EPA-M models are presented at anticipation window $h=\{640, 1280\}$~ms}
    \label{tab:op_points}
    \vspace{-10mm}
\end{table}

\subsubsection{Speculative Execution Strategy}

The EPA model runs as a parallel streaming module alongside
existing components, following a speculative execution logic:

\begin{enumerate}
    \item \textbf{Trigger \& Fork:} Upon anticipating an endpoint with horizon $h$, the system ``forks'' the conversation state. It triggers the LLM to generate a short look-ahead buffer (e.g., 10 tokens) based on the current partial transcript.
    
    \item \textbf{Pre-Synthesis (Cache):} These tokens are fed to the TTS engine to generate audio frames, which are stored in a \textit{speculative cache} rather than being played out.
    
    \item \textbf{Verification:} Wait for the duration of the horizon $h$.
    \begin{itemize}
        \item \textbf{Success (True Positive):} If the Unmute endpointer detects a turn completion within this window, the cached audio is released to the output buffer immediately, and generation continues by using the full transcript, and currently generated tokens. This effectively masks the processing latency.
        \item \textbf{Failure (False Positive):} If no endpoint is detected by the end of $h$, the anticipation module resumes. The cache is discarded when a new endpoint is anticipated.
    \end{itemize}
\end{enumerate}
When a speculative cache is available at the true endpoint, system latency is reduced to the response time of the endpointer alone, bypassing the ASR, LLM, and TTS generation bottlenecks.

\subsubsection{Evaluation Protocol}
We adopt the turn-taking evaluation framework from Full-Duplex Bench V1 \cite{lin2025full_v1}. We report two primary performance indicators: \textbf{Average Latency}, as defined by the benchmark\footnote{\url{https://github.com/DanielLin94144/Full-Duplex-Bench/}}, and our proposed \textbf{Expected Redundant Computation} (ERC). We omit metrics for other turn-taking capabilities (e.g., interruption handling), as our anticipation module is orthogonal to these functions and preserves the original performance of the Unmute framework.

\section{Results and Discussion}

\subsection{VAP vs EPA}
\begin{figure}[t]
    \centering  
    \includegraphics[width=1\linewidth]{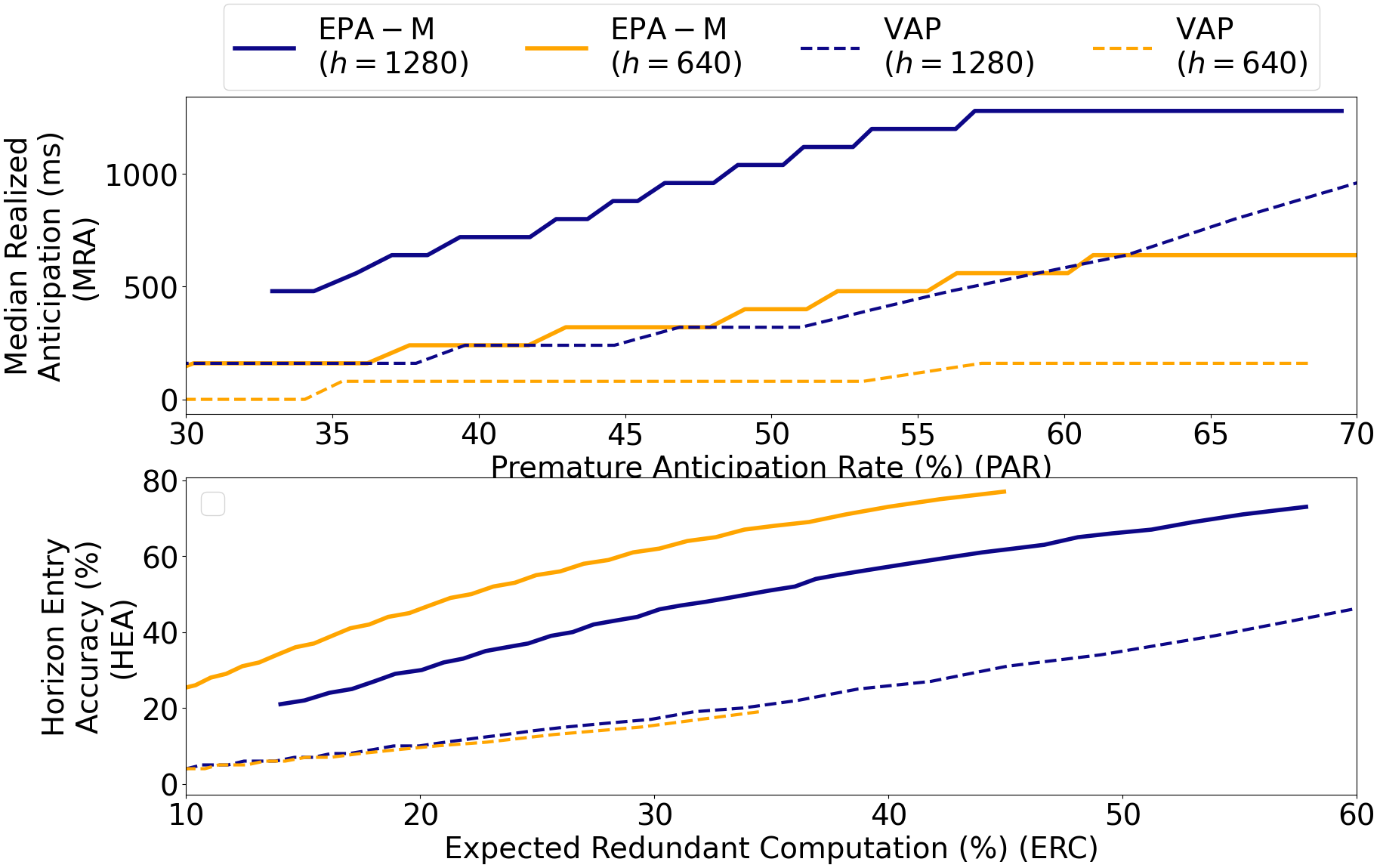}
    \vspace{-4mm}
    \caption{Comparison of the VAP baseline and the proposed EPA-M model at anticipation horizons, $h \in \{640, 1280\}$ ms, on SpokenWOZ test set}
    \label{fig:vap_vs_prop}
    \vspace{-4mm}
\end{figure}

Figure \ref{fig:vap_vs_prop} evaluates our proposed EPA-M model (Section \ref{sec:model_arch}) against the adapted VAP baseline. The results indicate a substantial gap in performance; EPA-M consistently dominates VAP across both trade-off spaces (MRA vs. PAR and HEA vs. ERC). While VAP's generalized training enables zero-shot adaptation to various turn-taking tasks, it struggles to produce precise, fixed-horizon endpoint forecasts. 

Furthermore, Table \ref{tab:op_points} explicitly compares these results across two Expected Redundant Computation (ERC) operating points of interest. For context, an ERC of 33\% implies that integration with downstream systems will result in 33\% redundant compute due to early, discarded predictions. At the $\approx 33\%$ ERC operating point, VAP and EPA-M yield comparable PAR. However, EPA-M achieves drastically higher HEA and MRA. For instance, at $h=640$~ms, EPA-M achieves a median anticipation (MRA) of 640 ms compared to VAP's 160 ms, showing that when a specific computational budget is allowed, EPA can reliably match the target anticipation horizon, whereas the VAP baseline falls significantly short. Additionally, at a stricter ERC constraint of $\approx 15\%$, EPA-M maintains an MRA of 480 ms with a 22.1\% Horizon Entry Accuracy (HEA), whereas VAP yields negligible anticipation capabilities.

\begin{figure}[b]
    \centering
    \vspace{-8mm}
    \includegraphics[width=1\linewidth]{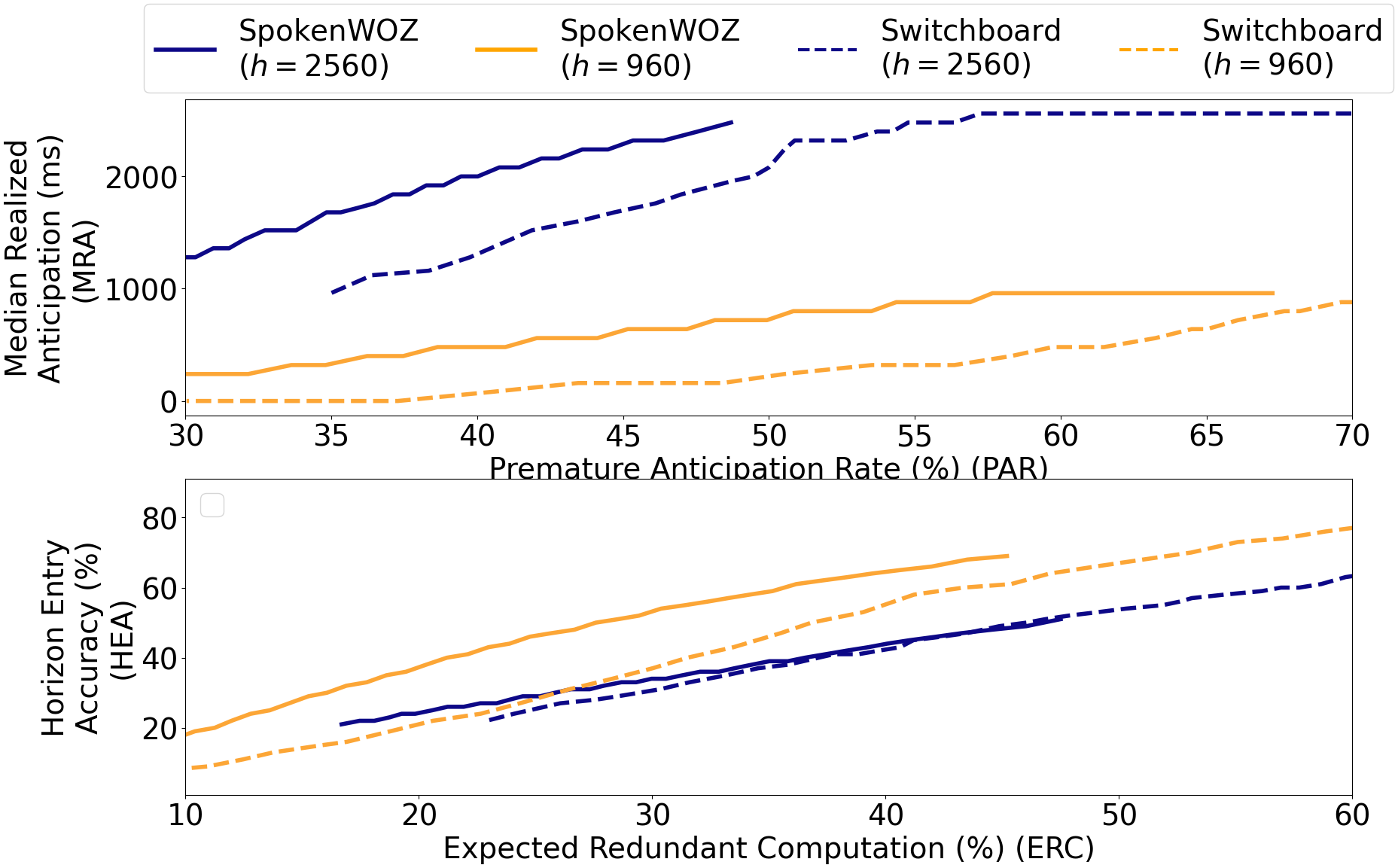}
    \vspace{-4mm}
    \caption{Performance across various anticipation horizons, $h \in \{960, 2560\}$ ms, for models trained on the SpokenWOZ (task-oriented) and Switchboard (conversational) datasets.}
    \label{fig:switch_spoke}
\end{figure}

% \begin{figure}[b]
%     \centering
%     \includegraphics[width=1\linewidth]{images/fcall_spokenwoz_ep_cutoff_vs_median_forecast.png}
%     \caption{Figure presents the performance different anticipation targets for STL, MHL and HCL}
%     \label{fig:diff_arch}
% \end{figure}

\subsection{Conversational vs. Task-Oriented Speech}
Figure \ref{fig:switch_spoke} compares performance on spontaneous (Switchboard Eval2000) and structured task-oriented (SpokenWOZ) dialogues. Across both short ($h=960$~ms) and long ($h=2560$~ms) horizons, the model achieves consistently superior anticipation on SpokenWOZ, yielding higher MRA for any fixed PAR, and higher HEA for any given ERC budget. This confirms that the inherent unpredictability of open-domain conversational speech poses a significant challenge, making endpoint anticipation currently most viable for structured applications.

\subsection{Endpoint Anticipation Architecture}
Table \ref{tab:op_points} compares the performance of our two proposed architectures: EPA-M and EPA-S (Section \ref{sec:model_arch}). We observe that both models show comparable results, displaying only minor deviations across target horizons and evaluation metrics. While this demonstrates that both architectural formulations are well-suited for endpoint anticipation, EPA-M can be flexibly deployed across varying target horizons without requiring horizon-specific retraining.

\subsection{System Integration Evaluation}
Table~\ref{tab:unmute_results} presents the latency evaluation of the Unmute system integrated with our EPA model (Section~\ref{sec:speech_speech_integration}). Results show that endpoint anticipation significantly reduces average system latency. We report a 505~ms reduction using local LLM and TTS components, these gains would likely be higher for API-based models with greater inherent latency. The residual $\approx$690~ms latency consists of standard execution on turns with low anticipation, semantic VAD trigger delays, and inter-process WebSocket communication. The 28.4\% ERC indicates that the volume of discarded speculative computation is restricted to roughly 28\% of the theoretical maximum for these turns. This overhead is lowered by efficient inference \cite{kwon2023efficient} and sharing the feature extractor for anticipation model and downstream tasks. Ultimately, this trade-off allows cascaded pipelines to support further real-time processing.

\begin{table}[t]
    \centering
    \begin{tabular}{l|c|c}
    \hline
         \textbf{System} & \textbf{Avg. Latency (ms) $\downarrow$} & \textbf{ERC (\%) $\downarrow$} \\
         \hline
         Unmute Baseline & 1195 & -- \\
         Unmute + EPA-M  & \textbf{690} & 28.4 \\
         \hline
    \end{tabular}
    \caption{Latency reduction and computational overhead of EPA-M ($h=960$~ms) integration within Unmute framework.}
    \label{tab:unmute_results}
    \vspace{-10mm}
\end{table}
\vspace{-3mm}
\section{Conclusion}
In this work, we introduced \textit{endpoint anticipation} as a framework to minimize turn-taking latency in modular spoken dialogue systems. By forecasting end-of-turn signals before speech completion, we enable speculative execution of downstream ASR, LLM, and TTS components. We explored two horizon modeling strategies---EPA-S and EPA-M---and proposed a set of metrics to quantify the practical trade-off between realized latency savings and computational redundancy. Our multi-target EPA-M model consistently outperforms competitive VAP-based baselines across both conversational and task-oriented datasets. 

Integrated into the \textit{Unmute} framework, our approach reduced average latency by 505~ms alongside a 28.4\% increase in speculative computation. This effectively masks sequential bottlenecks in cascaded architectures. We will open-source our implementation. Future work will explore semantic edge cases, such as mid-turn backtracking and late-arriving critical information.

\section{Generative AI Use Disclosure}
The authors used Gemini 3 Pro exclusively for language refinement. No AI tools were used to generate technical content. The authors assume full responsibility for this manuscript.

\bibliographystyle{IEEEtran}
\bibliography{mybib}

@article{likhomanenko2025chipchat,
  title={Chipchat: Low-latency cascaded conversational agent in mlx},
  author={Likhomanenko, Tatiana and Carlson, Luke and Bai, Richard He and Gu, Zijin and Tran, Han and Aldeneh, Zakaria and Zhang, Yizhe and Zhang, Ruixiang and Zheng, Huangjie and Jaitly, Navdeep},
  journal={arXiv preprint arXiv:2509.00078},
  year={2025}
}

@article{defossez2024moshi,
  title={Moshi: a speech-text foundation model for real-time dialogue},
  author={D{\'e}fossez, Alexandre and Mazar{\'e}, Laurent and Orsini, Manu and Royer, Am{\'e}lie and P{\'e}rez, Patrick and J{\'e}gou, Herv{\'e} and Grave, Edouard and Zeghidour, Neil},
  journal={arXiv preprint arXiv:2410.00037},
  year={2024}
}

@misc{roy2026personaplexvoicerolecontrol,
      title={PersonaPlex: Voice and Role Control for Full Duplex Conversational Speech Models}, 
      author={Rajarshi Roy and Jonathan Raiman and Sang-gil Lee and Teodor-Dumitru Ene and Robert Kirby and Sungwon Kim and Jaehyeon Kim and Bryan Catanzaro},
      year={2026},
      eprint={2602.06053},
      archivePrefix={arXiv},
      primaryClass={cs.CL},
      url={https://arxiv.org/abs/2602.06053}, 
}

@inproceedings{hu2025salm,
  title     = {{Efficient and Direct Duplex Modeling for Speech-to-Speech Language Model}},
  author    = {Ke Hu and Ehsan Hosseini-Asl and Chen Chen and Edresson Casanova and Subhankar Ghosh and Piotr Żelasko and Zhehuai Chen and Jason Li and Jagadeesh Balam and Boris Ginsburg},
  year      = {2025},
  booktitle = {{Interspeech}},
  pages     = {2715--2719},
  doi       = {10.21437/Interspeech.2025-874},
  issn      = {2958-1796},
}

@article{udupa2025streaming,
  title={Streaming endpointer for spoken dialogue using neural audio codecs and label-delayed training},
  author={Udupa, Sathvik and Watanabe, Shinji and Schwarz, Petr and Cernocky, Jan},
  journal={ASRU},
  url={arXiv:2506.07081},
  year={2025}
}

@article{inoue2024real,
  title={Real-time and continuous turn-taking prediction using voice activity projection},
  author={Inoue, Koji and Jiang, Bing'er and Ekstedt, Erik and Kawahara, Tatsuya and Skantze, Gabriel},
  journal={arXiv preprint arXiv:2401.04868},
  year={2024}
}

@inproceedings{chang17_interspeech,
  title     = {{Endpoint Detection Using Grid Long Short-Term Memory Networks for Streaming Speech Recognition}},
  author    = {Shuo-Yiin Chang and Bo Li and Tara N. Sainath and Gabor Simko and Carolina Parada},
  year      = {2017},
  booktitle = {{Interspeech}},
  pages     = {3812--3816},
  doi       = {10.21437/Interspeech.2017-284},
  issn      = {2958-1796},
}

@article{shih2025can,
  title={Can Speech LLMs Think while Listening?},
  author={Shih, Yi-Jen and Raj, Desh and Wu, Chunyang and Zhou, Wei and Bong, SK and Gaur, Yashesh and Mahadeokar, Jay and Kalinli, Ozlem and Seltzer, Mike},
  journal={arXiv preprint arXiv:2510.07497},
  year={2025}
}

@article{kuroki2025kame,
  title={KAME: Tandem Architecture for Enhancing Knowledge in Real-Time Speech-to-Speech Conversational AI},
  author={Kuroki, So and Kubo, Yotaro and Akiba, Takuya and Tang, Yujin},
  journal={arXiv preprint arXiv:2510.02327},
  year={2025}
}

@inproceedings{ekstedt2022voice,
  title     = {Voice Activity Projection: Self-supervised Learning of Turn-taking Events},
  author    = {Erik Ekstedt and Gabriel Skantze},
  year      = {2022},
  booktitle = {Interspeech},
  pages     = {5190--5194},
  doi       = {10.21437/Interspeech.2022-10955},
  issn      = {2958-1796},
}

@inproceedings{inoue-etal-2024-multilingual,
    title = "Multilingual Turn-taking Prediction Using Voice Activity Projection",
    author = "Inoue, Koji  and
      Jiang, Bing{'}er  and
      Ekstedt, Erik  and
      Kawahara, Tatsuya  and
      Skantze, Gabriel",
    editor = "Calzolari, Nicoletta  and
      Kan, Min-Yen  and
      Hoste, Veronique  and
      Lenci, Alessandro  and
      Sakti, Sakriani  and
      Xue, Nianwen",
    booktitle = "Proceedings of the 2024 Joint International Conference on Computational Linguistics, Language Resources and Evaluation (LREC-COLING 2024)",
    month = may,
    year = "2024",
    address = "Torino, Italia",
    publisher = "ELRA and ICCL",
    url = "https://aclanthology.org/2024.lrec-main.1036/",
    pages = "11873--11883"
}

@inproceedings{sakuma2023improving,
  title={Improving the response timing estimation for spoken dialogue systems by reducing the effect of speech recognition delay.},
  author={Sakuma, Jin and Fujie, Shinya and Zhao, Huaibo and Kobayashi, Tetsunori},
  booktitle={Interspeech},
  pages={2668--2672},
  year={2023}
}

@article{zink2024predictive,
  title={Predictive speech recognition and end-of-utterance detection towards spoken dialog systems},
  author={Zink, Oswald and Higuchi, Yosuke and Mullov, Carlos and Waibel, Alexander and Kobayashi, Tetsunori},
  journal={arXiv preprint arXiv:2409.19990},
  year={2024}
}

@inproceedings{sakuma2023response,
  title={Response timing estimation for spoken dialog systems based on syntactic completeness prediction},
  author={Sakuma, Jin and Fujie, Shinya and Kobayashi, Tetsunori},
  booktitle={2022 IEEE Spoken Language Technology Workshop (SLT)},
  pages={369--374},
  year={2023},
  organization={IEEE}
}

@article{arora2025stream,
  title={Stream rag: Instant and accurate spoken dialogue systems with streaming tool usage},
  author={Arora, Siddhant and Khan, Haidar and Sun, Kai and Dong, Xin Luna and Choudhary, Sajal and Moon, Seungwhan and Zhang, Xinyuan and Sagar, Adithya and Appini, Surya Teja and Patnaik, Kaushik and others},
  journal={arXiv preprint arXiv:2510.02044},
  year={2025}
}

@article{su2024roformer,
  title={Roformer: Enhanced transformer with rotary position embedding},
  author={Su, Jianlin and Ahmed, Murtadha and Lu, Yu and Pan, Shengfeng and Bo, Wen and Liu, Yunfeng},
  journal={Neurocomputing},
  volume={568},
  pages={127063},
  year={2024},
  publisher={Elsevier}
}

@article{zeghidour2025streaming,
  title={Streaming sequence-to-sequence learning with delayed streams modeling},
  author={Zeghidour, Neil and Kharitonov, Eugene and Orsini, Manu and Volhejn, V{\'a}clav and de Marmiesse, Gabriel and Grave, Edouard and P{\'e}rez, Patrick and Mazar{\'e}, Laurent and D{\'e}fossez, Alexandre},
  journal={arXiv preprint arXiv:2509.08753},
  year={2025}
}

@article{si2023spokenwoz,
  title={Spokenwoz: A large-scale speech-text benchmark for spoken task-oriented dialogue agents},
  author={Si, Shuzheng and Ma, Wentao and Gao, Haoyu and Wu, Yuchuan and Lin, Ting-En and Dai, Yinpei and Li, Hangyu and Yan, Rui and Huang, Fei and Li, Yongbin},
  journal={NeurIPS},
  volume={36},
  pages={39088--39118},
  year={2023}
}

@inproceedings{godfrey1992switchboard,
  title={SWITCHBOARD: Telephone speech corpus for research and development},
  author={Godfrey, John J and Holliman, Edward C and McDaniel, Jane},
  booktitle={ICASSP},
  volume={1},
  pages={517--520},
  year={1992},
  organization={IEEE}
}

@inproceedings{chang2020low,
  title={Low Latency Speech Recognition Using End-to-End Prefetching.},
  author={Chang, Shuo-Yiin and Li, Bo and Rybach, David and He, Yanzhang and Li, Wei and Sainath, Tara N and Strohman, Trevor},
  booktitle={Interspeech},
  pages={1962--1966},
  year={2020}
}

@misc{SileroVAD,
  author = {Silero Team},
  title = {Silero VAD: pre-trained enterprise-grade Voice Activity Detector (VAD), Number Detector and Language Classifier},
  year = {2024},
  publisher = {GitHub},
  journal = {GitHub repository},
  howpublished = {\url{https://github.com/snakers4/silero-vad}},
  commit = {insert_some_commit_here},
  email = {hello@silero.ai}
}

@article{stivers2009universals,
  title={Universals and cultural variation in turn-taking in conversation},
  author={Stivers, Tanya and Enfield, Nicholas J and Brown, Penelope and Englert, Christina and Hayashi, Makoto and Heinemann, Trine and Hoymann, Gertie and Rossano, Federico and De Ruiter, Jan Peter and Yoon, Kyung-Eun and others},
  journal={Proceedings of the National Academy of Sciences},
  volume={106},
  number={26},
  pages={10587--10592},
  year={2009},
  publisher={National Academy of Sciences}
}

@article{lin2025full_v1,
  title={Full-duplex-bench: A benchmark to evaluate full-duplex spoken dialogue models on turn-taking capabilities},
  author={Lin, Guan-Ting and Lian, Jiachen and Li, Tingle and Wang, Qirui and Anumanchipalli, Gopala and Liu, Alexander H and Lee, Hung-yi},
  journal={arXiv preprint arXiv:2503.04721},
  year={2025}
}

@article{vaswani2017attention,
  title={Attention is all you need},
  author={Vaswani, Ashish and Shazeer, Noam and Parmar, Niki and Uszkoreit, Jakob and Jones, Llion and Gomez, Aidan N and Kaiser, {\L}ukasz and Polosukhin, Illia},
  journal={NeurIPS},
  volume={30},
  year={2017}
}

@inproceedings{kwon2023efficient,
  title={Efficient Memory Management for Large Language Model Serving with PagedAttention},
  author={Woosuk Kwon and Zhuohan Li and Siyuan Zhuang and Ying Sheng and Lianmin Zheng and Cody Hao Yu and Joseph E. Gonzalez and Hao Zhang and Ion Stoica},
  booktitle={Proceedings of the ACM SIGOPS 29th Symposium on Operating Systems Principles},
  year={2023}
}

@inproceedings{arora25_interspeech,
  title     = {{Chain-of-Thought Training for Open E2E Spoken Dialogue Systems}},
  author    = {Siddhant Arora and Jinchuan Tian and Hayato Futami and Jee-weon Jung and Jiatong Shi and Yosuke Kashiwagi and Emiru Tsunoo and Shinji Watanabe},
  year      = {2025},
  booktitle = {{Interspeech}},
  pages     = {4833--4837},
  doi       = {10.21437/Interspeech.2025-2339},
  issn      = {2958-1796},
}

@article{arora2025landscape,
  title={On the landscape of spoken language models: A comprehensive survey},
  author={Arora, Siddhant and Chang, Kai-Wei and Chien, Chung-Ming and Peng, Yifan and Wu, Haibin and Adi, Yossi and Dupoux, Emmanuel and Lee, Hung-Yi and Livescu, Karen and Watanabe, Shinji},
  journal={arXiv preprint arXiv:2504.08528},
  year={2025}
}

@article{yu2025salmonn,
  title={Salmonn-omni: A standalone speech llm without codec injection for full-duplex conversation},
  author={Yu, Wenyi and Wang, Siyin and Yang, Xiaoyu and Chen, Xianzhao and Tian, Xiaohai and Zhang, Jun and Sun, Guangzhi and Lu, Lu and Wang, Yuxuan and Zhang, Chao},
  journal={NeurIPS},
  url={arXiv:2505.17060},
  year={2025}
}

@article{wang2025end,
  title={End-to-end Listen, Look, Speak and Act},
  author={Wang, Siyin and Yu, Wenyi and Chen, Xianzhao and Tian, Xiaohai and Zhang, Jun and Lu, Lu and Zhang, Chao},
  journal={ICLR},
  url={arXiv:2510.16756},
  year={2026}
}

@article{zufle2026f,
  title={F-Actor: Controllable Conversational Behaviour in Full-Duplex Models},
  author={Z{\"u}fle, Maike and Klejch, Ondrej and Sanders, Nicholas and Niehues, Jan and Birch, Alexandra and Lam, Tsz Kin},
  journal={arXiv preprint arXiv:2601.11329},
  year={2026}
}

@article{li2025easy,
  title={Easy Turn: Integrating Acoustic and Linguistic Modalities for Robust Turn-Taking in Full-Duplex Spoken Dialogue Systems},
  author={Li, Guojian and Wang, Chengyou and Xue, Hongfei and Wang, Shuiyuan and Gao, Dehui and Zhang, Zihan and Lin, Yuke and Li, Wenjie and Xiao, Longshuai and Fu, Zhonghua and others},
  journal={arXiv preprint arXiv:2509.23938},
  year={2025}
}

@article{zeng2024glm,
  title={Glm-4-voice: Towards intelligent and human-like end-to-end spoken chatbot},
  author={Zeng, Aohan and Du, Zhengxiao and Liu, Mingdao and Wang, Kedong and Jiang, Shengmin and Zhao, Lei and Dong, Yuxiao and Tang, Jie},
  journal={arXiv preprint arXiv:2412.02612},
  year={2024}
}

@article{levinson2015timing,
  title={Timing in turn-taking and its implications for processing models of language},
  author={Levinson, Stephen C and Torreira, Francisco},
  journal={Frontiers in psychology},
  volume={6},
  pages={136034},
  year={2015},
  publisher={Frontiers}
}

\end{document}